# Supercritical angle microscopy and spectroscopy

*Running title:* Fluorescence imaging from forbidden angles

Martin Oheim,*,a,✉ Adi Salomon,*,†,b and Maia Brunstein,*,‡

*Université de Paris, SPPIN - Saints-Pères Paris Institute for the Neurosciences, CNRS, Paris F-75006, France;
†Institute of Nanotechnology and Advanced Materials (BINA), Department of Chemistry, Bar-Ilan University, Ramat-Gan, 52900 Israel;
‡Chaire d'Excellence Junior, Université Sorbonne Paris Cité, Paris, F-75006 France.

Address all correspondence to

✉ [martin.oheim@parisdescartes.fr](mailto:martin.oheim@parisdescartes.fr)

SPPIN - Saints-Pères Paris Institute for the Neurosciences
Centre National de la Recherche Scientifique, CNRS UMR 8003
Université de Paris, Campus Saint-Germain-des-Prés
45 rue des Saints Pères
F-75006 Paris, France

Phone:  +33 1 42 86 42 21 (lab), -42 22 (office)

a) MO was a *Joseph Meyerhof Invited Professor* with the Department of Biomolecular Sciences, The Weizmann Institute for Science, Rehovot, Israel, during the academic year 2018-19.

b) Adi Salomon is currently a *Chateaubriand Senior Research Fellow* with the SPPIN.

ABSTRACT. Fluorescence detection, either involving propagating or near-field emission, is widely being used in spectroscopy, sensing and microscopy. Total internal reflection fluorescence (TIRF) confines fluorescence excitation by an evanescent (near-) field and it is a popular contrast generator for surface-selective fluorescence assays. Its emission equivalent, supercritical angle fluorescence (SAF) is comparably less established although it achieves a similar optical sectioning as does TIRF. SAF emerges when a fluorescing molecule is located very close to an interface and its near-field emission couples to the higher-refractive index medium ($n_2 > n_1$) and becomes propagative. Then, most fluorescence is detectable on the side of the higher-index substrate and a large fraction of this fluorescence is emitted into angles forbidden by Snell's law. SAF as well as the undercritical angle fluorescence (UAF) (far-field emission) components can be collected with microscope objectives having a high-enough detection aperture $NA > n_2$ and be separated in the back-focal plane (BFP) by Fourier filtering. The BFP image encodes information about the fluorophore radiation pattern, and it can be analysed to yield precise information about the refractive index in which the emitters are embedded, their nanometric distance from the interface and their orientation. A SAF microscope can retrieve this near-field information through wide-field optics in a spatially resolved manner, and this functionality can be added to any existing inverted microscope.

[221 words]

**Significance statement**

Many processes happen close to or at interface. Battery electrodes, biofilm growth, transport across cellular membranes, or phase mixing are just a few examples of such phenomena. While fluorescence-based detection techniques are standard, their surface-selective variants require the elimination of bulk background fluorescence to detect near-interface molecules in isolation. Near-field techniques usually require specialized instrumentation, but there is an optical phenomenon called supercritical-angle fluorescence (SAF) that makes near-field information available in the far field. SAF can originate from near-interface fluorophores only, and these molecules emit a highly directional fluorescence. SAF can be captured and imaged in the back-focal plane of an objective lens and it contains not only information about the molecule, but also about its environment.

{MAIN TEXT}

## I. SUPERCRITICAL ANGLE FLUORESCENCE, A TWIN AND COMPANION TO TIRF

*"If you can see me, I can see you"*

OPTICAL RECIPROCITY[1] is key for understanding SUPERCRITICAL ANGLE FLUORESCENCE (SAF), or "forbidden-light detection" (1). SAF has been dubbed "evanescence in emission" (2), much as TOTAL INTERNAL REFLECTION fluorescence (TIRF) refers to "evanescence in excitation". Common to either optical phenomena are near-field interactions of light with a dielectric material. When light hits the interface, the angles of refraction and reflection are governed by SNELL'S LAW, and the transmitted and refracted intensities are given by the Fresnel coefficients, **Fig. 1***A* (*top*). For supercritical angles, $\theta_2 \geq \theta_c = \mathrm{asin}(n_1/n_2)$, the reflection is total (i.e., $I_{\mathrm{refl}} = I_0$) and an evanescent wave (EW) is generated just above the boundary as a consequence of energy and momentum conservation. This inhomogeneous surface wave skims the optically rarer medium and it propagates along $+x$ for an incidence from the left as illustrated in panel A. The EW intensity decays exponentially with distance $z$ from the interface with a length-constant $\delta = \lambda/[4\pi \cdot (n_2^2 \cdot \sin^2(\theta) - n_1^2)^{1/2}]$ of the order of $\lambda/4$. Here $\lambda$ denotes the wavelength of light. If a fluorophore is present at a height $h$ above the interface, then the EW excites fluorescence in proportion to $I(h)/I_0 \sim \exp[-h/\delta(\theta)]$, **Fig. 1***A* (*bottom*). Thus, fluorescence excitation will be confined to a thin slice of $\approx 2\delta$ thickness (see, e.g., (3, 4)).

Fluorescence emission has both near- and far-field components. The far-field of a radiating dipole displays the well-known 'laying-eight' angular emission pattern with an intensity maximum perpendicular to the dipole axis, **Fig. 1***B* (*top*). When the molecule rotates, e.g., in solution its time-averaged far-field emission will be isotropic. For a dipole located at a height $h \gg \lambda$ from the dielectric interface the collected fraction of isotropically emitted fluorescence scales as $\Omega/4\pi = \Phi = \frac{1}{2}[1 - \cos(\theta_{NA})] = \frac{1}{2}[1-(1-(NA/n_2)^2)^{1/2}]$, where $\theta_{NA} = \operatorname{asin}(NA/n_2)$. For a NA 1.45 TIRF objective, $\Phi$ is of the order of 36%, i.e., only one third of the emitted fluorescence is collected despite the large NA of the TIRF objective lens.

---

[1] Terms in SMALL CAPS are explained in the Glossary at the end of this Perspective Article.

In addition to its far-field, dipole emission has an evanescent component, too. Normally unperceived in far-field detection[2], this near-field can generate light detectable at supercritical angles when the dipole is located very close to the dielectric boundary ($h \leq \lambda$). The evanescent emission is converted by the proximity of the dielectric bounder into a propagating wave, similar to the appearance of a propagating, transmitted wave when a second prism is positioned in close proximity above the reflecting interface in the case of frustrated TIR (5). Analogous to the classical 'double-prism' experiment, this propagating wave is directed into angles "forbidden" by Snell's law, and the phenomenon is hence termed "supercritical angle fluorescence" (SAF), **Fig. 1***B* (*bottom*) (2, 6, 7). Only fluorophores located very close to the dielectric boundary can emit SAF. The closer they are the more of their emission is funneled into supercritical angles making SAF. Its near-exponential distance dependence makes SAF an absolute nanometric axial ruler insensitive to most aberration biases (6, 8-10). In turn, fluorophores distant from the interface cannot emit SAF at all, and their fluorescence is undercritical, **Fig. 1***C*. Thus, the selective detection of SAF vs. undercritical angle fluorescence (UAF) allows discriminating near-interface fluorophores against those located in the bulk (1).

> {*PULL OUT* TEXT} In TIRF, light propagating at high angles produces a near-field close to an interface; in SAF, a near-field close to an interface generates high-angle emission. This is the optical reciprocity of TIRF and SAF.

**II. SELECTIVE SAF DETECTION PERMITS THE ACQUISITION OF TIRF-LIKE IMAGES**

---

[2] Because either the detection NA is too small or the signal emitted from near-interface dipoles is overwhelmed by the dominant emission from bulk fluorophores.

Equally based on near-field/interface interactions, SAF and TIRF produce a similar optical sectioning. SAF collection in the solid-angle range above $\theta_c$ renders fluorescence *detection* surface-selective, much as supercritical fluorescence *excitation* confines TIRF. {PULL-OUT TEXT CLOSE TO HERE} The similar axial confinement of both near-field techniques is illustrated in a cell-biological application by the pseudo-color overlay of TIRF and SAF images of green-fluorescent protein- (GFP-) expressing mitochondria in cultured cortical astrocytes taken with the same 1.46-NA objective, **Fig. 2**A (*i*). TIRF images were acquired upon azimuthal beam spinning (11) at polar beam angles of $\theta$ = 64, 68 and 70°, respectively, (*green*), and the SAF image was taken upon EPI-excitation ($\theta$ = 0, *red*). For TIRF, the precise incidence angle $\theta$ determines the EW penetration depth $\delta(\theta)$ (although the 'true' excitation light distribution can be contaminated by unwanted stray light and long-range excitation components (4, 12)). On the contrary, SAF exclusively depends on the axial fluorophore height *h*, so that the apparent overlap between TIRF and SAF images will generally vary with $\theta$ (13). Using Pearson's and Manders' coefficients (14) to quantify, respectively, the amount of co-variance and overlap between the 'red' and 'green' pixel intensities, we recognize that TIRF and SAF highlight similar structures ($R_{12} \approx 0.8$) but the fraction of pixels that overlap on the segmented SAF image with TIRF pixels ($M_2$) is steadily diminishing for higher $\theta$, **Fig. 2**B. This is plausible, because the TIRF excitation volume shrinks with increasing angle and becomes gradually more selective, eventually leading to a thinner optical section than SAF at very high beam angles, **Fig. 2**A (*ii*).

However, grazing beam angles require positioning the focused laser spot very close to the objective's limiting $\theta_{NA}$, which can produce stray excitation and degrade the excitation confinement and image contrast (12). In objective-type TIRF (15), several studies have documented an unwanted 5-10% of non-evanescent (long-range) excitation components that adds to the localized EW excitation (4, 12, 16, 17). Advantageously, TIRF excitation and SAF detection can be associated to relieve constraints on $\theta$ (13). Due to the combined excitation confinement and emission axial filtering, TIR-SAF images display a better optical sectioning than TIRF alone, as result of suppressing the unwanted fluorescence from deeper sample regions. Yet, as only the supercritical part of dipole radiation is being used, SAF images tend to be dimmer but – as a consequence of the efficient background suppression – they have a better signal-to-background ratio (SBR) than TIRF-alone images for the same excitation intensities as seen in a proof-of-concept experiment using tiny fluorescent microspheres as a standard, **Fig. 2**C.

Of course, SAF does not require TIRF excitation and SAF detection has been combined with many other excitation geometries, including epi-fluorescence, confocal-spot excitation without (1, 18, 19) and with stimulated-emission depletion (STED) (20). A growing body of spot-excitation and confocal SAF detection is devoted to supercritical fluorescence correlation spectroscopy (FCS) (21-24). Finally, dipole-surface interactions and supercritical-angle detection are not restricted to fluorescence: supercritical-angle luminescence (SAL) detection (25) and supercritical-angle RAMAN (SAR) spectroscopy (26) have been reported.

## III. SELECTIVE DETECTION OF SUPER- CRITICAL EMISSION COMPONENTS

Whereas TIRF can be detected without specialized optics[3], **Fig. 3**A, for SAF to be detected one requires, (*i*), a *detection* NA>$n_1$ spanning supercritical angles and, (*ii*), a device for angular emission filtering to reject the undercritical emission component that is dominated by fluorophores in the bulk volume.

When using highly corrected imaging optics, e.g., a high-NA microscope objective, Abbe's sine condition defines an unambiguous relationship between emission angles and radii in the objective's backfocal plane (BFP), providing a rationale for FOURIER-PLANE FILTERING, $r_\theta = f_{obj} \cdot n_2 \cdot \sin\theta$. Here, $f_{obj}$ is the objective's focal length, $f_{TL}/M$, i.e., the ratio of the focal length of the manufacturer tube lens and the objective transversal magnification; see Ref. (27) for the subtleties of Abbe's sine condition for very-high NA objectives. One might imagine to block UAF emission by simply placing an appropriately sized opaque disk in the BFP. However, the BFP is located inside the objective lens, **Fig. 3***B*, so that a relay telescope is needed to image the BFP into a CONJUGATE PLANE, **Fig. 3***C*. A caveat is that in either of the geometries shown in panels *B* and *C* the effective detection aperture is reduced to a ring, so that the lower-NA Fourier components are clipped, and only a tiny part of the objectives SPATIAL-FREQUENCY BANDWIDTH is used for light collection and image formation, resulting, respectively, in poor sensitivity and a degraded point-spread function. Thus, the SAF image obtained by just blocking out UAF has a lower resolution than the corresponding image taken at full aperture (28, 29). The precise positioning of an aperture stop is facilitated by a BERTRAND LENS, **Fig. 3***D*.

---

[3] But TIRF needs a high-NA objective ref. (15) or and external prism ('Kretschmann geometry') for guiding light a supercritical angles to the interface.

The spatial-resolution problem is elegantly avoided with *virtual SAF* (*v*SAF) detection (30). Here, an iris rather than a central disk is used as an aperture mask, and two images are acquired: one taken at the full NA, and the other with the effective detection aperture stopped down to NA' ≈ RI$_{sample}$ ($\Leftrightarrow r_c = f_{obj} \cdot n_1$, where $r_c$ is the critical radius marking the transition of UAF to SAF), **Fig. 3***E*. The first image contains both SAF and UAF emission components, the latter contains UAF only. Their difference image, *v*SAF = (SAF + UAF) – UAF, is a 'virtual' SAF image, equivalent to the one taken with the ring aperture and containing exclusively the fluorescence components emitted into supercritical angles. However, as two high-NA images were used for image formation, the lateral resolution is maintained on the *v*SAF image, which is an important feature for microscopic imaging.

Alternative geometries do not use commercial objectives at all. In early SAF work, a custom high-NA parabolic ring-mirror (1) surrounding a central undercritical detection (and excitation) optical path was used instead of a high-NA objective. The same authors later presented a custom objective design that integrates separate high-NA and low-NA pathways (22). We ignore if this objective is commercially available. Common to both geometries is that they use a point detection and require sample scanning for SAF-image acquisition (18), similar to the formally equivalent confocal spot-scanning TIRF geometry (31-33).

## IV. SAF APPLICATIONS: MORE THAN SURFACE-SELECTIVE FLUORESCENCE

Depending on whether simple aperture filtering is used or the radiation pattern is imaged and the BFP image analyzed, SAF has different applications.

### SAF applications with aperture filtering

*Near-surface fluorescence.* Surface-selective fluorescence detection is straightforward for non-imaging (sensing) applications (34, 35) or confocal-spot geometries (1, 18, 22). SAF is increasingly being used for surface binding assays, the characterization of antibodies or sensitive toxin detection. In an attempt of scaling up SAF sensing, miniaturized parabolic optics have been integrated into the bottom of disposable plastic biochips (36) and polymer test tubes (37). By further size reduction, a truncated cone-shape geometry was shown to parallelize sensing in 256 spots on a microscope-slide sized biochip (38, 39). Useful for sensing, the large aberrations and missing central cone in these devices are prohibitive for imaging applications.

*A short-sighted microscope.* Implemented in a scanning geometry combining confocal-spot illumination with SAF collection on a point detector (18, 33), as a full-field technique either with a resolution-compromising aperture disk (29) or in the resolution-preserving *v*SAF variant (13, 30, 40), SAF microscopy is a simple, cheap and less alignment-demanding alternative to TIRF, in which no laser source is needed. Compatible with a host of illumination schemes and readily implemented on a standard inverted microscope, *v*SAF imaging is increasingly being used for imaging cell adhesion, cytoskeletal, vesicle- and membrane dynamics. Used in conjunction with TIRF (13) or its structured-illumination variant TIRF-SIM (41-45), SAF can help to remove impurities, imperfections and unknowns associated with objective-type EW excitation (4, 40) and provide an isotropic ∼100-nm resolution.

*Nanometer-axial fluorophore localization.* The steep dependence of the SAF fraction of emitted fluorescence on the axial fluorophore distance *h* has prompted the use of SAF as a nanometric axial ruler with single-molecule sensitivity in a 150-200 nm range close to the interface (7, 10, 22). 3-D single-molecule localization with nm precision is offered through the combination of dSTORM and SAF (8). Hence, one prominent use of SAF is to remove the traditional imbalance of lateral and axial resolution in super-resolution microscopies by the SAF-based axial single-molecule localization. Applications include PALM, STORM, STED and TIRF-SIM microscopies with combined SAF detection.

*Fluorescence correlation spectroscopy (FCS).* The combined effect of confocal detection and SAF-mediated surface confinement results in a tiny (attoliter, $10^{-18}$ l) detection volume. It has been used for ultrasensitive near-interface diffusion and concentration measurements (21, 24, 46). Simultaneous UAF and SAF detection permits the synchronous readout of bulk and near-surface mobility and concentration (47) and removes ambiguities resulting from axial motion in non-flat membrane stretches (23).

**SAF applications with BFP imaging and the analysis of the radiation pattern**

SAF goes beyond axial optical sectioning when imaging the aperture (pupil) plane of a high-NA objective rather than its field (sample) plane, **Fig. 3***D*. The BFP image contains the fluorophore radiation pattern, which contains fluorophore information (see (10) for a recent overview). Depending on the exact geometry used, this information is derived either from a limited number of fluorophores in the confocal spot (33) or else integrated (averaged) over the entire field of view (40). How the aberrations most relevant to Fourier microscopy, including alignment tolerances, apodization, the effect of magnification on the modulation transfer function, and vignetting-induced reductions of the effective NA for wide-field measurements influence Fourier-plane images has been studied in detail (27).

*Imaging molecular orientation.* Albeit not called SAF at the time but "Fourier-plane image analysis", the dependence of the fluorophore radiation pattern on the orientation of the dipole axis relative to the optical axis has allowed studying molecular orientation (48), rotational dynamics (49, 50) and even nanometric axial movement in single-molecule (9) and single-nanoparticle detection assays (8), using a combination of lateral single-molecule localization and SAF-based axial-distance assays. Back focal imaging and spectroscopy have been applied not only to study angular emission patterns of fluorescent molecules, but also to investigate anisotropic Raman signals from molecules, elastic scattering, plasmonic scattering and non-linear scattering as well as secondary emission via optical antennas (see, e.g., (51) for a recent review).

*Variable-angle SAF spectroscopy.* Surface-selective (immuno-) assays often require a wash step prior to read-out due to the otherwise overwhelming signal of the large number of unbound (bulk) fluorescent molecules that dominate over the signal from the molecules of interest, bound to the substrate. TIRF has been the method-of-choice for investigating the population of surface-bound fluorophores in isolation, and variable-angle scans (VA-TIRF) have allowed a 'tomographic' reconstruction of layer thickness, membrane topography, or cell adhesion from fluorophore distance (52-58). SAF is an obvious alternative, but the angular dependence of SAF on both the surface and bulk fluorescence contributions had not been experimentally studied until recently (59). The van Dorpe lab studied how the collection angle influences the SAF surface sensitivity in the presence of bulk fluorophores. Two different fluorophores were used. One was bound to the surface and the other was dissolved in the bulk. They measured the spectrum at discrete points in the back focal plane (BFP) and quantified the relative contribution of the two fluorophores by spectral unmixing. While the highest signal-to-noise ratio was observed in the region just above $\theta_c$ because of the higher signal intensity, the highest signal-to-bulk ratio was observed at much larger angles >68° for a glass/water interface. Thus, for experiments where bulk exclusion is important, increasing the angle of SAF collection enhances the surface sensitivity but at the cost of a decreased signal intensity. This study inspires studies combining emission spectral imaging and spectral BFP image analysis for multi-color studies of the basal cell membrane or the nanoscale 3-D organization, e.g., for understanding the nano-architecture of cell adhesion sites and the peripheral multilayered cytoskeleton, which – at present - are typically being studied with variable-angle TIRF. Also, understanding the angular dependence on the sensitivity of a SAF biosensor will allow tuning the collection angles towards specific applications and contribute to designing smaller, more efficient devices.

*Verification of high-NA objectives.* When the RI of the fluorophore-embedding medium is known (as for air, water, or calibrated sucrose[4] or glycerol/water solutions (60, 61)), the calibration of BFP radii via $r_c = f_{obj} \cdot n_1$ on a BFP image and application of Abbe's sine condition allows measuring the outer, bounding radius. This measurement directly provides the effective objective $NA_{eff}$ via $r_{NA} = f_{obj} \cdot n_2 \cdot \sin\theta = f \cdot NA_{eff}$. In this manner, large NAs (beyond $n_2 \cdot \sin\theta_c$) that are otherwise difficult to measure because of TIR can be controlled experimentally by imaging the angular radiation pattern of surface-bound fluorophores (62). Building on this strategy, we found important deviations between the true and specified NA for some objectives (Fig. S2 in (12)), a relevant information prior to buying such expensive objective lenses.

*SAF refractometry.* Conversely, with $NA_{eff}$ known, an interesting and comparably new SAF application is near-surface refractometry (40): as the transition of UAF to SAF depends on the sample RI $n_1$ (but not on $n_2$), the measurement of the "critical radius" $r_c = f_{obj} \cdot n_1$ on the BFP image allows for a sensitive measurement of the *local* RI of the fluororphore-embedding medium. The precision of this measurement depends on the quality of the fit of the BFP image (and hence on the available signal, pixel size, and camera sensitivity), and typical RI values with ±0.001 error are being reached. This property has been used – together with diverse fluorophores that we subcellularly targeted to different organelles – for intracellular RI-measurements (40). Of note, even though wide-field SAF refractometry measures the *local* fluorophore environment, it reports an *average* RI value, $\langle n_1 \rangle$, averaged over all SAF-emitting fluorophores in the field-of-view. In an attempt to achieve spatially resolved RI measurements, a combination of Bessel-beam-spot excitation with confocal intensity and simultaneous SAF-based RI detection through the same high-NA objective has been reported in a recent communication (33).

---

[4] Degrees Brix (°Bx) is the sugar content of an aqueous solution. One °Bx is 1 g of sucrose in 100 g of solution and represents the strength of the solution as mass %. The °Bx is traditionally used in the wine, sugar, carbonated beverage, fruit juice, maple syrup and honey industries, but the commercial availability of calibrated sucrose solution has also been exploited in science. See, e.g., Ref. 40.     Brunstein, M., L. Roy, and M. Oheim. 2017. Near-membrane refractometry using supercritical angle fluorescence. Biophys. J. 112(9):1940-1948..

*Label-free RI-based SAF assay.* An interesting variant of this refractometric readout is SAF detection from a fluorophore-coated glass microcapillary (63). Here, the analyte is non-fluorescent and the RI change induced by a medium change or - as shown by the authors - bacterial growth on the capillary bottom, is detected as a shift in $r_c$ and hence $<n_1>$. In the same manner, one can imagine to measure cell adhesion or confluency in unlabeled cell cultures grown on a fluorophore-coated coverslip. In either case, the sensitivity of this label-free fluorescent SAF assay will be limited by the fluorescence signal available after prolonged use of the capillary sensor and by the timecourse of photobleaching of the immobilised dye. Conceptually, the technique is reminiscent of nm-scale axial resolution with non-radiative excitation, (64).

*Incidence-angle calibration in VA-TIRF.* BFP imaging of the SAF-ring has been used for calibrating $\theta$ in objective-type TIRF microscopy. Similar to the earlier RI-measurement, the approach is based on a series of $\theta_c$-measurements in the objective BFP using solvents with different RIs. In one study, a thin homogenous layer of fluorophores in air was used to produce a SAF ring of known radius $r_{c(air)} = f_{obj}$ that is compared to the annular distribution of excitation light in azimuthal beam-spinning TIRF (65). Two follow-up studies used the same strategy with a thin layer of organic dye for TIRF-SIM calibration (45), or for VA-TIRF using Qdots embedded in PMMA (66). In the latter case, the fluorophore layer was topped with solutions of different, known RI (measured with an Abbe refractometer) and the laser spot (seen when the emission filter was taken out) steered to these known angles to attain a more precise multi-angle calibration.

**V. WHAT'S NEXT?**

SAF, either used for surface-selective fluorescence detection or – combined with an analysis of BFP images – for imaging the fluorophore radiation pattern, is an active and rapidly expanding area of research. Beyond its early uses in spectroscopy, sensing and studies of single-molecule mobility and rotation, SAF is increasingly being used as an imaging technique, either with wide-field detection or else as a point-measurement technique with sample scanning. Recent studies emphasized the quantitative analysis of BFP images to extract various fluorophore properties and ongoing work aims at the real-time analysis of BFP images through rapid feature extraction. Future studies will likely combine of SAF-based BFP-image variable-emission angle analysis with fluorescence-intensity, life-time or anisotropy-based contrast, marking the arrival SAF-based contrast in multi-modal microscopies.

Given that these applications all rely on BFP image analysis, we expect deep-learning to have an impact similar to other fields of microscopic image analysis. Artificial intelligence (AI) based BFP image analysis will boost information-extraction from the narrow high-intensity ring characteristic for SAF. Fast feature recognition – particularly when multiple emitters or multi-color emitters are present - and adaptive aperture filtering are only two aspects that will benefit from AI, in a manner similar to exploiting the chromatic dependence of PSF shapes from different emitters imaged on a grayscale camera (67).

The relative ease of use, reduced complexity and (compared to TIRF) low price of instrumentation required for *v*SAF is most likely to increase its dissemination of the for routine biological microscopy. A '*v*SAF-module' that can be fitted between the microscope exit port and an existing EMCCD or sCMOS camera is now commercially available (by the French start-up Abbelight – non-related to the authors), and makes the technique accessible for non-expert users.

Finally, in as much as SAF does not depend on how fluorescence was excited but represents an intrinsic fluorophore property depending solely on its distance from the interface, we foresee a growing use of SAF in microscope calibration and standardization applications (4).

**List of abbreviations**

| | | |
|---|---|---|
| AI | - | artificial intelligence |
| BFP | - | back-focal plane |
| EMCCD | - | electron-multiplying charge-coupled device |
| EW | - | evanescent wave |
| FCS | - | fluorescence correlation spectroscopy |
| FWHM | - | full width at half maximum |
| NA | - | numerical aperture |
| PCC | - | Pearson's correlation coefficient |
| PMMA | - | poly(methyl methacrylate) |
| PSF | - | point spread function |
| RI | - | refractive index |
| SAF | - | supercritical angle fluorescence |
| SBR | - | signal-to-background ratio |
| sCMOS | - | scientific complementary metal oxide sensor |
| STED | - | stimulated emission depletion microscopy |
| TIR(F) | - | total internal reflection (fluorescence) |
| UAF | - | under-critical angle fluorescence |

**Glossary**

APERTURE (OR BACKFOCAL) PLANE – in a multi-lens system, a plane conjugate to the aperture, in a light microscope, the set of conjugate aperture planes comprises the exit pupil, objective back-focal plane, condenser front focal plane and the lamp filament. (see also field plane for the other set of conjugate planes including the sample plane and detector).

BERTRAND LENS – an optical device used in aligning the various optical components of a light microscope. It allows observation of the back focal plane of the objective lens and its conjugated focal planes.

FOURIER FILTERING – refers to modulating or clipping in frequency space; in SAF, it refers to shaping the SPATIAL FREQUENCY BANDWIDTH of the microscope by selectively attenuating, blocking or enhancing certain segments of the (conjugate) back-focal plane by an aperture mask.

NUMERICAL APERTURE (NA = $n \cdot \sin\theta_{NA}$) – Magnitude related to the maximal angle of collection or excitation of an optical element.

(OBJECTIVE) BACK-FOCAL PLANE (BFP) – Effective rear focal plane of the assembly of lens composing the objective. For most objective lenses, it is located inside the objective and hence only accessible by generating an intermediate image in a conjugate plane.

OPTICAL RECIPROCITY – (for all practical grounds here) principle that describes how a ray of light and its reverse follow the same optical path.

RAMAN (SCATTERING) – Inelastic scattering of a photon when it encounters matter.

SNELL'S LAW – A law from geometrical optics that describes how a ray of light is reflected and/or transmitted when it encounters an interface between two different refractive-index media, $n_i \cdot \sin\theta_i = const$.

SPATIAL-FRENQUENCY BANDWIDTH - Range of frequencies that is transmitted by the system. In an optical (or imaging) system this bandwidth is inversely related to the spatial optical resolution. See also Object transfer function and Point Spread Function.

SUPERCRITICAL ANGLE FLUORESCENCE (SAF) – Fluorescence that propagates trough a higher-index medium (compared to the emission medium) at 'forbidden' angles higher than the critical angle.

TOTAL INTERNAL REFLECTION (TIR) – Complete bouncing back of a beam from a transparent interface (i.e., no transmitted light) that can occur when a light impinges at an interface from the higher- ($n_2$) to the lower-refractive index side ($n_1$), at an angle larger than a certain critical angle, $\theta_c = \mathrm{asin}(n_2/n_1)$.

ACKNOWLEDGEMENTS

This study was supported by a *Chaire d'Excellence Junior University Sorbonne Paris Cité*, USPC (to MB), the Israeli Science Foundation (ISF-NSFC 2525/17, to AS), the European Union (H2020 Eureka! Eurostars NANOSCALE (to MO and AS). Our collaborative research on TIRF and SAF was further financed by a French-Israeli ImagiNano CNRS-LIA grant (to AS & MO), the *Agence Nationale de la Recherche* (ANR-10-INSB-04-01, *grands investissements* FranceBioImaging, FBI, to MO), the Université Paris Descartes (invited professorship during the academic year 2017-18, to AS) and the French *Ministère des Affaires Etrangères* (Chateaubriand Senior Fellowship to AS). The authors thank Dana Khanafer for help with Fig. 2*C*. The Oheim lab is a member of the C'nano IdF and *Ecole de Neurosciences de Paris* (ENP) excellence clusters for nanobiotechnology and neurosciences, respectively.

**Author contributions**

All authors contributed to experiments. The manuscript was written by MO with contributions of all authors. All authors have given their approval to the final version of the manuscript.

**Notes**

The authors declare no competing financial interest.

**FIGURES**

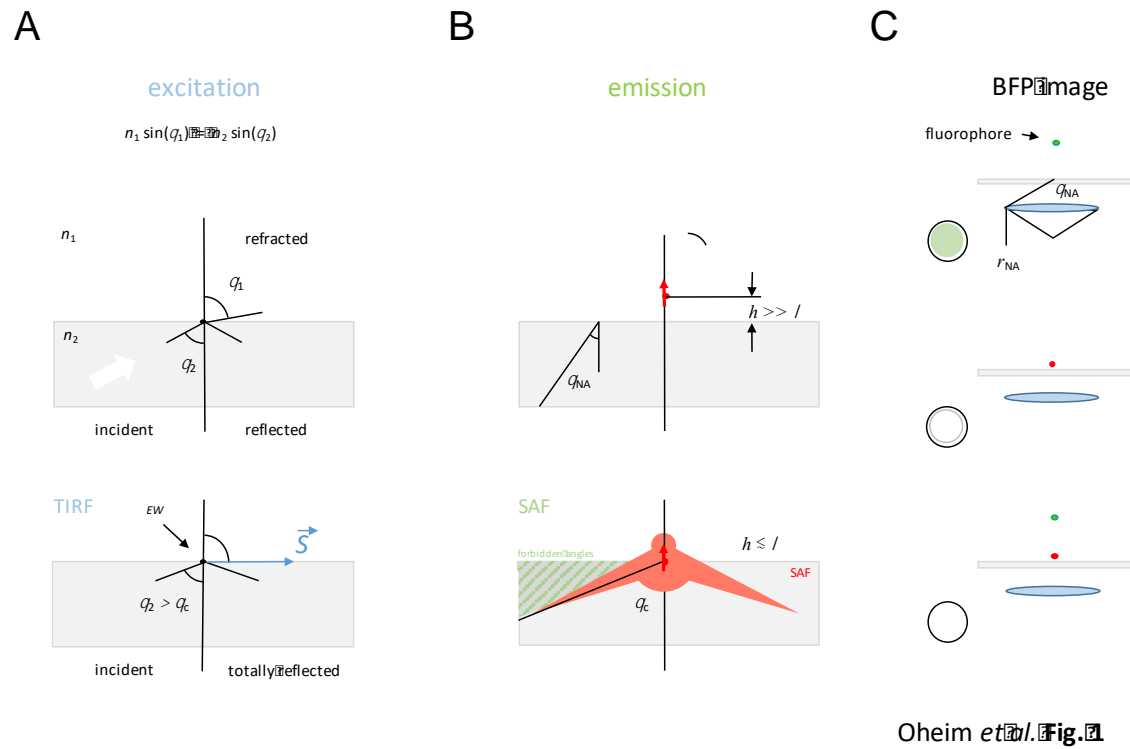

Oheim *et al.* **Fig. 1**

FIGURE 1. *Evanescence in excitation and emission.* (A), *top*, we assume for the purpose of this paper a planar dielectric interface separating two homogenous media having different refractive indices (RIs), $n_2 > n_1$. Snell's law $n_1 \cdot sin(\theta_1) = n_2 \cdot sin(\theta_1)$ governs the angles of refraction at a dielectric interface. For a glass/water interface ($n_2 > n_1$), light impinging at the interface from the side of the denser medium 2 is totally reflected beyond a critical angle of incidence, $\theta_c = \mathrm{asin}(n_2/n_1)$. Total reflection creates an inhomogeneous, propagating evanescent wave (EW) right above the dielectric boundary in medium 1, *bottom*. **S** is the Pointing vector, indicating the direction of energy flux. (B), fluorescence emission of near-interface dipoles. Far-field emission pattern from a radiating dipole located far from the interface (fluorophore height $h \gg \lambda$) showing the familiar "laying-8" pattern. Molecular rotation (arrow) in solution often results in an isotropic fluorescence emission (dashed circle). The collected fraction of fluorescence hence depends on the solid angle captured by the NUMERICAL APERTURE (NA) of the objective. On the contrary, for $h \lesssim \lambda$, fluorescent specimen does not emit fluorescence isotropically, but approximately 2/3 of the fluorescence is emitted into the higher-index medium. Of this emission, the main part is directed to solid angles above the critical angle. On the contrary, already for $h = \lambda/2$, fluorescence above the critical angle is decreased dramatically. Hence, like TIRF, supercritical angle fluorescence (SAF) is suited to discriminate between molecules at or near to surfaces and other fluorophores in the bulk. (C), provided the detection NA of the collection optics is sufficiently large (NA > $n_1$), supercritical (SAF) and undercritical fluorescence emission components (UAF) can be collected simultaneously and separated in the in the objective backfocal plane (BFP). SAF and UAF can be distinguished as they travel at different NAs (i.e., radii in the BFP), which can be used for aperture filtering (see main text).

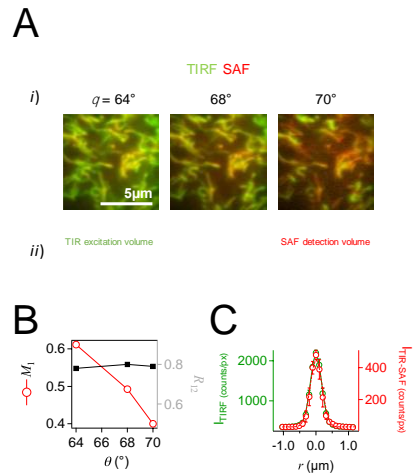

Oheim *et al.* **Fig. 2**

FIGURE 2. *SAF detection of TIRF-like images and combined TIR-SAF acquisition with improved contrast.* (A), TIRF (*green*) and (epi-excited) SAF (*red*) images of EGFP-expressing mitochondria in a cultured cortical astrocyte. For TIRF the polar beam angle was $\theta$ = 64, 68 and 70° as indicated. (B), Evolution of Manders' co-localization coefficient $M_1$ for SAF (image 1, *red*) and of Pearsons correlation coefficient (PCC, *black*, see Ref. (14)). TIRF and SAF highlight similar structures ($<R_{12}>$ = 0.79 ± 0.01). Unlike PCC, $M_1$ measures the amount of TIRF pixels coinciding with SAF pixels following thresholding of each channel. TIRF and SAF images are similar for intermediate beam angles (lower EW penetration depths), before TIRF becomes more discriminative at very high $\theta$. However, this object-based analysis obscures the gain in sensitivity by SAF. (C), Contrast of TIRF-alone and combined TIR-SAF images. Images of yellow-green emitting 93-nm diameter beads upon 488-nm excitation and the FWHM and Michelson contrast ('visibility') calculated as $C_M$ = ($I_{max}$ - $I_{min}$)/($I_{max}$ + $I_{min}$). Pixel size was 103 nm, $t_{exp}$ = 50 ms. Symbols are measured intensity mean ± *SD* from four independent measurements. Through lines are weighted Gaussian fits. SBR was 4.92 ± 0.16, $C_M$ = 0.71 and FWHM = 293 ± 2 nm for TIRF alone; vs. SBR = 14.4 ± 1.1, $C_M$ = 0.88, FWHM = 311 ± 1 nm for TIR-SAF. Thus, while no large difference was observed between FWHMs, the combined optical sectioning of TIRF excitation and SAF detection almost tripled contrast (×2.93).

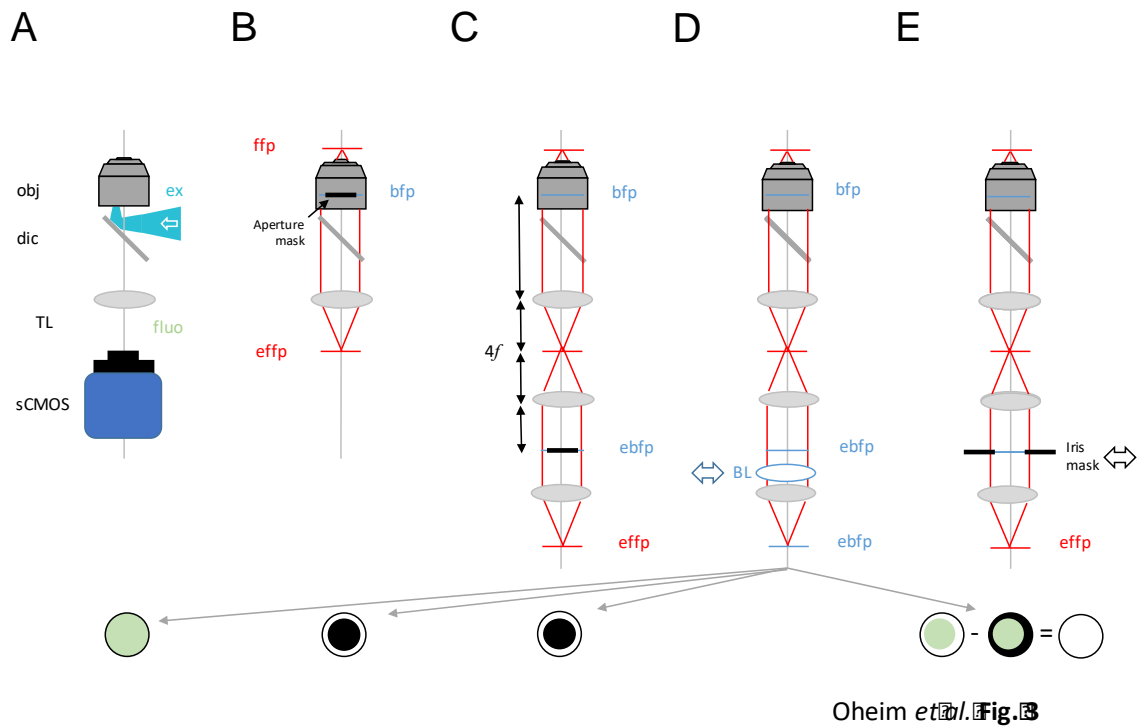

Oheim *et al.* **Fig. 3**

FIGURE 3. *Microscope geometries for TIRF and SAF detection.* (A), typical optical layout for prismless ('through-the-objective') type evanescent-wave excitation. In the simplest case, an expanded laser beam is focused in an eccentric position in the BFP of a high-NA objective. ex – excitation light, obj – objective lens, dic – dichroic mirror, TL – tube lens, sCMOS – scientific complementrary metal-oxyde sensor. (B), hypothetical selective SAF detection with an opaque aperture mask in the objective BFP. (C), pracitical implementation of SAF detection with an aperture mask placed in a conjugate aperture plane, produced by a '4*f*' relay telescope. Like the arrangement shown in panel C, this scheme results in a resolution loss due to the ring aperture and clipping of central rays. (D), insertion fo a Bertrand lens (BL) allows BFP (or "Fourier-plane") imaging. (E), virtual SAF, *v*SAF configuarartion, in which an iris stop is used instead of an opaque disk and the SAF image is calculated from the difference of two images taken with the iris open and stopped down to *NA'* = $n_1$, respectively.